\newcommand{\beq} {\begin{eqnarray}}
\newcommand{\eeq} {\end{eqnarray}}
\documentclass[12pt, preprint,preprintnumbers,amsfonts,aps,nofootinbib]{revtex4}
\usepackage{graphicx}
\begin{document}

\title{Worldlines as Wilson Lines}

\author{Daniel Green}
\email{drgreen@stanford.edu}
\affiliation{SLAC and Department of Physics, Stanford University, Stanford, CA 94305-4060} 

\preprint{SU-ITP-08/09}
\preprint{SLAC-PUB-13223}

\begin{abstract}
Gravitational theories do not admit gauge invariant local operators.  We study the limits under which there exists a quasi-local description for a class of non-local gravitational observables where a sum over worldlines plays the role of the Wilson line for gauge theory observables.  We study non-local corrections to the local description and circumstances where these corrections become large.  We find that these operators are quasi-local in flat space and AdS, but fail to be quasi-local in de Sitter space.
\end{abstract}

\maketitle
\section{Introduction}
One of the striking features of gravitational physics is that the effective field theory description seems to fail in circumstances where, naively, it should be valid.  The best studied example is the black hole information paradox \cite{Hawking:1976ra}.  The loss of information can be derived entirely in a regime where the curvature is small \cite{Lowe:1995ac}.  Thus, if the evolution is unitary, it would seem to require a failure of effective field theory \cite{Susskind:1993if,Hooft:1990fr}.  Recently, studies of de Sitter space have suggested that eternal de Sitter may be another example where effective field theory fails to give sensible predictions \cite{Banks:2002wr,Goheer:2002vf, ArkaniHamed:2007ky, Giddings:2007nu}.

However, in theories of gravity, it is unclear what one should be calculating \cite{Giddings:2005id,Rovelli:1990ph,Marolf:1994wh}.  Local operators are not gauge invariant and there is no obvious non-local observable to replace them.  Therefore, one has no definitive measure of when the local operator description is a reliable approximation.  As a result, one could ask whether the apparent failure of effective field theory is simply a result of using poorly defined quantities.  

These issues do not arise in the absence of gravitational fluctuations.  On a fixed manifold, one can construct exactly local, diffeomorphism invariant operators.  For example, one can introduce Stueckelberg fields that make local observables manifestly diffeomorphism invariant \cite{ArkaniHamed:2007ky}.  We also suspect that, when gravity can be described perturbatively, local operators give the leading behavior of some non-local diffeomorphism invariant observable.  In principal, if these observables were known, one could compute the non-local corrections.

One can ask the same question in more conventional gauge theories.  For example, a charged field $\Psi(x)$ in a $U(1)$ gauge theory transforms non-trivially under local gauge transformations.  The requirement that observables are invariant under the gauge symmetry should not make the electron unphysical.  This intuition is correct and quasi-local gauge invariant observables are well known.  In particular, by attaching a Wilson line to the operator we get a gauge invariant observable, $\mathcal{O}(x) = \Psi(x) \exp(i e \int^{\infty}_{x} A_{\mu} dx^{\mu})$.

This example provides two important lessons.  First of all, when perturbation theory is valid, $\Psi(x)$ is a good approximation to the observable ($\mathcal{O}(x) \simeq \Psi(x) + \mathcal{O}(e)$), with the Wilson line giving rise to corrections that are perturbative in $e$.  Furthermore, if we turned off the dynamics of the gauge field (sending $e \to 0$), the exact observable is $\Psi(x)$.  Secondly, the Wilson line now gives us a precise measure of when $\Psi(x)$ is a good observable.  In particular, an IR divergent contribution from the Wilson line is the signal of confinement (area law for the Wilson loop).

In this paper, we will study the origin and form of the corrections to local correlation functions in gravitational theories.  We begin by discussing several variants of `worldline observables.'  These will be manifestly gauge invariant operators where a sum over worldlines plays a similar role to the Wilson line in gauge theory.  These operators are local to leading order (when perturbation theory is valid) but have corrections that are perturbative in the gravitational coupling.  We then discuss the implications to the black hole information paradox and to de Sitter space.  We find that the local description fails in eternal de Sitter.  Finally, we discuss the consequences for more general backgrounds.

\section{Observables}

In a certain sense, the only difference between observables in general relativity and gauge theory is simply that in gauge theory the appropriate observables are known.  While it has been well established that there are no gauge invariant local operators in theories of gravity, there has been little agreement on what should replace them.  There have been many proposals for gauge invariant operators (e.g. \cite{Page:1983uc,Rovelli:1990ph,Rovelli:1990pi,Tsamis:1989yu,Rovelli:2001bz, Giddings:2005id,Marolf:1994wh,Gambini:2004pe,Dittrich:2005kc}) but it remains unclear which, if any, are the ``correct'' observables.  For this reason, it may be helpful to use gauge theory as a guide when possible.

In gauge theory, for a field $\Psi_{a}(x)$ transforming in the fundamental representation, there are simple quasi-local observables constructed from Wilson lines.  In particular, in flat space the operator $ \mathcal{P}\exp(-\int_{x C}^{\infty}A_{\mu}dx^{\mu})^{a}_{b} \Psi_a (x)$ is gauge invariant \footnote{On a compact space, the Wilson line must end on a field in the anti-fundamental representation.  This is the familiar fact that the net charge must vanish on a compact space.}.  The gauge invariance arises from two features we will try to emulate in the gravitational context.  First, the trace over the index $a$ cancels the transformations of the Wilson line against the transformation of the operator.  The second feature is that we are only dividing out by the gauge transformations such that $g(y) \to 1$ as $y \to \infty$.  Transformations without this property represent global symmetries.

If we were to make the analogy between gravity and gauge theory, the integral $\int d^dx \sqrt{-g}$ acting on a local scalar operator, $\mathcal{O}(x)$, behaves like the trace over the gauge index.  Unfortunately, integrated operators alone completely wash away local information.  Therefore, we will need another ingredient if we are to get quasi-local behavior in some limit.

One approach is to add additional fields to the theory with spatially varying vacuum expectation values ($Z^{\mu} \simeq x^{\mu} +\delta Z^{\mu}$) \cite{Giddings:2005id,Gary:2006mw, Marolf:1994wh}.  We could then localize the integral over the manifold onto points where these fields take specific values ($\int d^dx \sqrt{-g} \mathcal{O}(x)\delta^{d}(Z^{\mu}-\lambda^{\mu})$).  This approach seems problematic for several reasons.  Primarily, it is surprising that one would need to change the field content to define observables.  For example, our experience with gauge theory suggests that adding new forms of matter changes the phase structure \cite{Fradkin:1978dv,Banks:1979fi}; thus, it is not clear that one would even be describing the same dynamics.  More pragmatically, we make measurements all the time without reference to a spatially varying scalar field.

Another approach is to use single particle states to localize operators \cite{Giddings:2005id,Rovelli:1990ph,Rovelli:1990pi}.  We will follow this approach as it seems more consistent with our intuition.  Our experience with measurement is based entirely on single and many particle states.  Furthermore, this is the type of approach that is successful in gauge theory.  In particular, one can think of the vev of the Wilson loop as the response of the system to an extremely massive quark-antiquark pair.  Nevertheless, these quarks are not really present in the dynamics of the theory and do not change the vacuum structure.

Before introducing the operators of interest, we should establish what we are looking for in an observable. (1)  It should be manifestly gauge invariant.  (2)  It should be exactly local when dynamical gravity is absent without adding new matter content to the theory.  (3)  In the presence of perturbative gravity, the operator should be quasi-local with computable gravitational corrections.  Here quasi-local means that, given a particular observable and a choice of gauge, the observable is well approximated by a local operator at a point or small region in coordinate space.  The operators discussed in the next subsection will have all these properties.

\subsection{Worldline Operators}

We will study observables in gravitational theories with specific asymptotia.  In particular, we will assume that we are modding out by diffeomorphisms that are trivial asymptotically.  Defining local observables in such backgrounds is sufficiently confusing that understanding this simpler problem may be helpful in understanding more general backgrounds.
  
For spaces with asymptotia, we can specify a point on the interior using information at the boundary.  The general idea is to use operators of the form
\beq
\label{ob1}
\mathcal{O}_{\Psi_{i=1..N}} = \int d^d x \sqrt{-g} \mathcal{O}(x) W(x, \Psi_{1},..,\Psi_{N}),
\eeq
where $\mathcal{O}(x)$ is a local scalar operator and $W(x,\Psi_{i})$ is some operator whose value at $x$ depends on boundary data $\Psi_{i}$.  In order to get local data from (\ref{ob1}), $W(x,\Psi_{i})$ should be a sharpely peaked function of $x$.  If $W(x,\Psi_{i})$ transforms under small diffeomorphisms as a scalar operator of $x$ then (\ref{ob1}) is invariant.  For operators of this type, one wants to define $W(x,\Psi_{i})$ such that $\mathcal{O}_{\Psi_{i=1..N}}$ is invariant under small diffeomorphisms and
\beq
\mathcal{O}_{\Psi_{i=1..N}} \simeq \mathcal{O}(\tilde{x}),
\eeq
where the point $\tilde{x}$ is defined by the boundary data.

A location in spacetime can be defined by the point where the wordlines of multiple particles intersect.  Classically, this point is specified exactly given boundary conditions on the paths of the particles.  Quantum mechanically, these trajectories are no longer exact and the resolution of points is smeared.  Nevertheless, the classical limit can be recovered by taking infinitely massive particles.  When we include gravity, this limit causes infinite backreaction.  For such a model, this is the origin of the breakdown of locality.

The basic observables we will study are built from path integrals over worldlines.  These single particle probes can be used to localize the operators onto the intersection of geodesics.  Specifically, we will consider observables where $W(x,\Psi_{i})$ is a product of several worldline path integrals.  Much like the gauge theory example, these need not be single particle states of some field added to the theory.

We will work with observables such that, in the classical limit, the operator is localized at the intersection of geodesics.  This can be achieved with operators of the form
\beq
\label{ob2}
\mathcal{O}_{\Psi_{i=1..N}} = \mathcal{N} \int d^d x \sqrt{-g} \mathcal{O}(x) \prod_{i=1}^{N} \int \mathcal{D}e(\tau) \int^{Y_{i}^{\mu}(1)=x^{\mu}}_{Y_{i}^{\mu}(0)= \Psi_{i}^{\mu}}  \mathcal{D}Y_{i}^{\mu}(\tau) e^{-\int d\tau ((\alpha' e)^{-1}\dot{Y}_{i}^{\nu}\dot{Y}_{i}^{\mu}G_{\mu \nu}+ \alpha' e m_{i}^{2})}  ,
\eeq
where $e$ is the worldline metric and $\alpha'$ is a constant with units of space-time length squared.  Here $\mathcal{O}(x)$ is some scalar operator, $m$ can be interpreted as the mass of a particle whose worldline is $Y^{\mu}(\tau)$ and $\mathcal{N}$ is a normalization constant.  These operators include $N$ worldlines starting in some asymptotic states $\Psi_{i}$ but ending at the same point on the interior, $x^{\mu}$.  This end point is where the scalar operator $\mathcal{O}$ is located.  This observable is diffeormorphism invariant for two reasons.  First, we integrate $x^{\mu}$ over the entire manifold. Second, the diffeomorphisms being modded out are assumed to leave the asymptotic states $\Psi^{\mu}_{i}$ unchanged.

\begin{figure}
\includegraphics[width=0.8\textwidth]{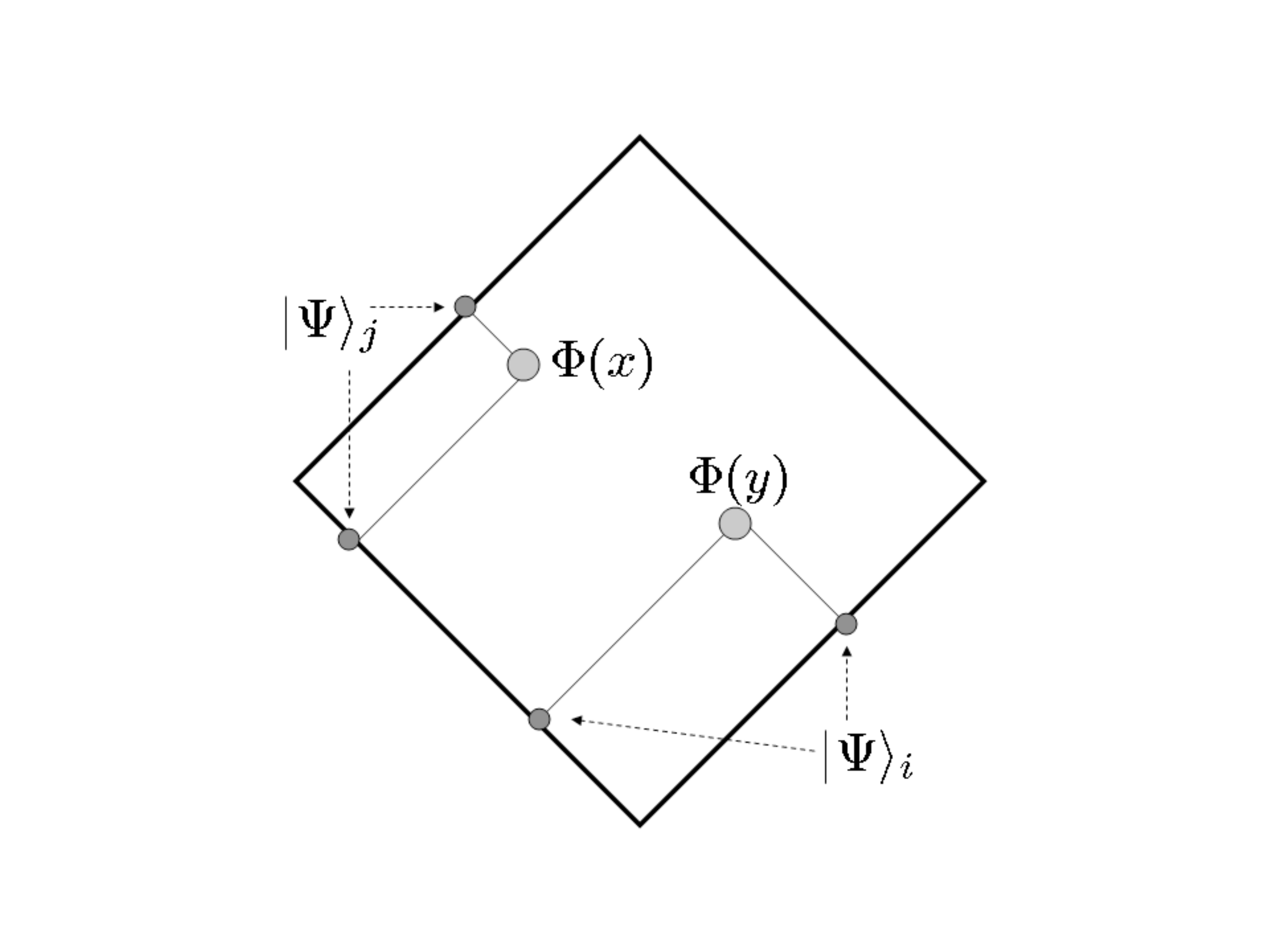}
\caption{\label{fig1} The operator $\Phi(z)$ is localized at the intersection of several null geodesics using (\ref{ob2}).  The choice of asymptotic states $|\Psi \rangle_{i}$ or $|\Psi \rangle_{j}$ determines whether $\Phi$ is localized around the point $y$ or $x$ respectively.}
\end{figure}

The final integral over $x$ will localize the operator onto the overlaps of the different worldlines.  The product of different worldlines will suppress contributions to the final $x$ integral away from the overlap of the wavefunctions.  Thus, operators localized at different points arise only from different choices of the asymptotic states (see figure \ref{fig1}).  Observables of this type are similar to the $\psi^2\phi$ model considered in \cite{Giddings:2005id}.  The limits $\alpha' \to 0$ or $m \to \infty$ play the roll of the classical limit for relativistic and non-relativistic particles respectively.  Therefore, for appropriate choices of the initial states, these limits applied to (\ref{ob2}) will give exactly local operators ($\mathcal{O}(x)$) localized at the intersection of geodesics.  

In the absence of dynamical gravity, there is no obstacle to taking the $\alpha' \to 0$ or the $m \to \infty$ limits.  The worldlines do not influence the dynamics of the other fields and thus we can take such limits without any trouble.  This is consistent with our expectation: observables should reduce to local operators in the absence of a fluctuating metric.

When we turn on gravitational fluctuations, localization fails as expected.  These worldlines couple to the metric as can be seen directly in (\ref{ob2}).  The coupling of gravity to these worldlines is either $\alpha'^{-2} M_{pl}^{-2}$ or $m^2 M_{Pl}^{-2}$ depending on the limit.  As a result, the gravitational corrections to our observables would be uncontrollably large in the classical limit.  Thus, $m$ and $\alpha'$ must be finite in a gravitational theory. 

Corrections to local operators in gravitational theories will appear in two ways.  The first type of correction arises from basic quantum mechanical effects at finite $m$ and $\alpha'$.  The non-locality here arises because of the spread of the wavefunction for the particles of interest.  The second type of correction comes from gravitational couplings to the worldlines and are perturbative in $G \equiv M^{-2}_{Pl}$.  This is consistent with gauge theory, where the Wilson line also produces perturbative, non-local corrections.  However, just as in gauge theory, this is not a signal of a fundamental breakdown in the locality of the theory.  For example, the S-matrix may be consistent with locality to all orders in perturbation theory \cite{Gross:1987ar,Mende:1989wt,Veneziano:2004er,Giddings:2007qq}.

The techniques for calculating the gravitational corrections for non-relativistic particles are well known \cite{Goldberger:2004jt,Goldberger:2005cd, Goldberger:2007hy}.  If we work with energies that are small compared to $M_{Pl}$, we can break the metric into a background and fluctuations, $G_{\mu \nu} = \tilde{G}_{\mu \nu}+ M^{-1}_{Pl} h_{\mu \nu}$.  For a non-relativistic particle worldline, the couplings of the worldline ($X^\mu$) to the graviton ($h_{\mu \nu}$) in a flat background are of the form
\beq
\label{coupling}
S_{coupling} =  \frac{m}{2 M_{Pl}} \int d{\tilde{\tau}} h_{\mu \nu} {X'}^{\mu} {X'}^{\nu} + \frac{m}{8 M^{2}_{Pl}} \int d{\tilde{\tau}} (h_{\mu \nu} {X'}^{\mu} {X'}^{\nu})^{2}+ \ldots
\eeq
where $\tilde{\tau}$ is the proper time along the geodesic and $X' \equiv \frac{\partial X}{\partial \tilde{\tau}}$.  Along with the usual Feynman rules for the perturbative treatment of gravity, one can use these couplings to compute the gravitational effects on the worldlines.  When the background includes two worldlines, graviton exchange between the particles must be included.  The ladder diagrams for this process can be resummed to get an effective potential for $X^{\mu}(\tau)$ \cite{Goldberger:2004jt,Goldberger:2005cd,Goldberger:2007hy}.  Unfortunately, we will find that non-relativistic worldlines do not give good local observables.  Specifically, the dispersion of the wavefunction causes the geodesic approximation to break down in finite time.  Therefore, we will look at the corrections in a relativistic example.

\subsection{Flat Space Example}

We will consider the massless limit of (\ref{ob2}) for perturbative fluctuations around a flat metric.  The definition of each worldline path integral (on a fixed metric) is well known \cite{Cohen:1985sm,Strassler:1992zr}, and after gauge fixing can be written as
\beq
\int \mathcal{D}e \mathcal{D}Y^{\mu} e^{-S}= \int_{0}^{\infty} dc c^{-\frac{d}{2}} e^{-\frac{(\Delta x)^2}{2 \alpha' c}},
\eeq
where $\Delta x^{\mu} = x^{\mu}_{i} - x^{\mu}$ is the difference between the fixed initial and final points.  We can now apply this to (\ref{ob2}) with $d=4$,
\beq
\mathcal{O}_{\Psi_{i=1..N}} = \mathcal{N} \int d^4 x \mathcal{O}(x) \prod_{i=1}^{N} \int_{\delta}^{\infty} dc_{i} c_{i}^{-2}e^{-\frac{(\Delta x_{i})^2}{2\alpha' c_{i}}}+\mathcal{O}(M_{Pl}^{-2})
\eeq
where $\mathcal{O}(M_{Pl}^{-2})$ are perturbative gravitational corrections.  We will determine the form of these corrections after determining the leading behavior.  We added $\delta$ to regulate the integral near $c \to 0$ as this corresponds to large energies and momenta which will couple strongly to gravity.  By a change of variables $a_{i} = c^{-1}_{i}$, these integrals take a more familiar form
\beq
\mathcal{O}_{\Psi_{i=1..N}} = \mathcal{N} \int d^4 x \mathcal{O}(x) \prod_{i=1}^{N} \int^{\frac{1}{\delta}}_0 da_{i} e^{-a_{i}\frac{(\Delta x_{i})^2}{2\alpha'}}+\mathcal{O}(M_{Pl}^{-2}).
\eeq
This integral is poorly defined if we work in Lorentzian signature.  We could resolve this by working with a Euclidean metric or Wick rotating the worldline time $\tau$ to get an imaginary action \footnote{We will ignore possible subtleties with this procedure.  Null geodesics are special to Lorentzian manifolds.  Thus, it is unclear how one should rigorously define the path integral while maintaining these special features.}.  For either choice, it is clear that performing the $a_{i}$ integral gives a sharply peaked function $\alpha' (\Delta x)^{-2}(1-e^{-\frac{(\Delta x)^2}{2  \alpha' \delta}})$.  The width of these peaks is controlled by $\alpha' \delta \equiv \Lambda^2$.  Note that in Lorentzian signature this is peaked on a null surface defined by $(\Delta x)^2 =0$ as opposed to a point in Euclidean signature.  The leading behavior is given by
\beq
\label{delta}
\mathcal{O}_{\Psi_{i=1..N}} \simeq \int d^4 x \mathcal{O}_{\Lambda}(x) \prod_{i=1}^{N} \delta(\Delta x_{i}^2) + \mathcal{O}(\Lambda^2 M_{Pl}^{-2}, \Lambda^{-1}),
\eeq
where $\mathcal{O}_{\Lambda}(x)$ the operator integrated over a region of size $\Lambda^{-1}$ centered at $x$ and we have chosen $\mathcal{N}$ to remove normalization constants from (\ref{delta}).  Thus, for each worldline, we get an approximate delta function which is smeared on scales smaller than $\Lambda^{-1}$.  With four worldlines, we get approximate localization in four dimensional flat space.

Thus far we have not used asymptotic flatness to make our observables gauge invariant.  In asymptotically flat space, we take the diffeomorphisms to unity at  $\mathcal{I}^{\pm}$.  Therefore, to make these observable gauge invariant we need to take the limit where the initial point ($x^{\mu}_{i}$) is taken to $\mathcal{I}^{\pm}$.  Working with $\tau \in (-\infty,0]$ makes this slightly easier but there is no obstruction to taking this limit.

Now, let us compute the gravitational corrections to leading order in $G = M_{Pl}^{-2}$.  The metric fluctuations couple to the worldline through the interaction
\beq
S_{coupling} = \frac{1}{M_{Pl}} \int d{\tau} (\frac{1}{\alpha' e} h_{\mu \nu} {\dot{X}}^{\mu} {\dot{X}}^{\nu}).
\eeq
The leading correction comes from a single graviton exchange between any two worldlines, $a$ and $b$.  Expanding in $M_{Pl}^{-1}$, the leading gravitational contribution is
\beq
\label{corr}
\mathcal{O}(M_{Pl}^{-2}) \simeq  \frac{1}{M_{Pl}^{2}} \int d^4x \langle  \mathcal{O}(x) \rangle \sum_{a,b=1}^{N} \int d \tau_{a} \int d \tau_{b}  \langle  \frac{1}{c_{a} c_{b} \alpha'^2} \dot{X}_{a}^{\mu} \dot{X}_{a}^{\nu} \mathcal{P}_{\mu \mu \lambda \rho}(X_{a},X_{b}) \dot{X}_{b}^{\lambda} \dot{X}_{b}^{\rho} \rangle _{x} ,
\eeq
where
\beq
\mathcal{P}_{\mu \nu \lambda \rho}(X_{a},X_{b}) = \int d^d k \frac{1}{2}(\eta_{\mu \lambda} \eta_{\nu \rho}+\eta_{\mu \rho} \eta_{\nu \lambda}-\frac{2}{d-2}\eta_{\mu \nu} \eta_{\lambda \rho})\frac{i}{k^2+i\epsilon}
\eeq
is the Feynman propagator for the graviton in De Donger gauge and $\langle \rangle_{x}$ is the expectation value, including the worldline path integrals with endpoints at $x$.  The Feynman rules for gravity are well known \cite{Hooft:1974bx, Bern:2002kj} and can be used to calculate the contributions at any order.  For the leading contribution, we will evaluate $\dot{X}$ on the classical solution and take $c_{i} \simeq \delta_{i}$.  Because the classical solutions are null ($\dot{X}^{\mu} \dot{X}_{\mu} = 0$), these terms vanish when $a=b$.  In a full calculation, one would include scattering of the worldlines due to the graviton exchange.  By requiring that the gravitational corrections are small for the classical solutions, we will also ensure that this is a subleading effect.

In order to understand our corrections, we should rewrite them in terms of physical quantities in spacetime.  The first thing to note is that $\Lambda^2 \dot{X}^{\mu} \equiv p^{\mu}$ has units of spacetime momentum.  Thus, we can rewrite (\ref{corr}) as 
\beq
\label{corr2}
\mathcal{O}(M_{Pl}^{-2}) \simeq  \langle \mathcal{O} (\tilde{x}) \rangle  \sum_{a \neq b}\frac{(p_{a}^{\mu}p_{b\mu})^2}{\Lambda^4 M_{Pl}^2} \int d \tau_{a} \int d \tau_{b} \int \frac{d^4 k}{(2\pi)^{4}} \frac{e^{i k_{\nu}(X^{\nu}_{a}(
\tau_{a})-X^{\nu}_{b}(\tau_{b}))}}{k^2}.
\eeq
where $\tilde{x}$ is a point defined by the intersection of null geodesics.  Because our operators are local to the scale $\delta X \sim \Lambda^{-1}$, we should work with energies of the order $(p_{a}^{\mu}p_{b\mu}) \sim k^2 \sim \Lambda^2$.  Therefore, we get a contribution of the form
\beq
\mathcal{O}(M_{Pl}^{-2}) \simeq \langle \mathcal{O}(\tilde{x})\rangle \frac{N(N-1)}{2} \frac{\Lambda^2}{M_{Pl}^2}.
\eeq
Therefore, for $\Lambda \ll M_{Pl}$ (for small $N$) we get an quasi-local operator to leading order with small gravitational corrections.  This is precisely what one would expect from the uncertainty principal.  We will see that unexpected complications can arise in different backgrounds.

\subsection{UV Completion}

General Relativity is a well defined, effective field theory below the scale $M_{Pl}$.  The problem of defining observables in this effective theory would seem to be a separate problem from the ultra-violet (UV) behavior of the theory.  Nevertheless, it is a natural question as to how any such proposal would fit into a UV completion.

Although the observables described above were motivated by effective field theory and gauge theory considerations, they seem to fit nicely with observables in AdS-CFT (for a review, see \cite{Aharony:1999ti}).  Many observables in the CFT appear in the bulk as propagators or minimal area surfaces that are fixed on the boundary.  The leading behavior is determined by the classical solutions, with higher order effects arising in the same way as above.  There have also been many attempts to define local observables in terms of the CFT \cite{Balasubramanian:1998sn,Banks:1998dd, Balasubramanian:1999ri}, some of which are similar to our proposal \cite{Bena:1999jv,Hamilton:2005ju}.

The probes in AdS-CFT are part of the fundamental particle content of the theory \footnote{Entanglement entropy is an exception, as it is also related to a minimal area surface \cite{Ryu:2006bv} but not one that arises from some more fundamental degree of freedom.}.  The masses of these particles can be quite large, arising from KK modes or solitonic objects (D-branes).  As a result, these probes can have masses above the cutoff scale of our low energy effective field theory.  From the low energy point of view, these massive probes will not be involved in the dynamics of the theory.  Thus, our observables seem consistent with the observables in a UV completion.

AdS-CFT suggests a broader class of observables that may be useful in gravitational theories.  Many operators in the CFT are described by minimal area surfaces of various dimensions.  We will focus on the worldline in this paper, but it may useful to consider integrals over other submanifolds as a tool for studying effective field theories with gravity.

\section{Limits of Locality}
\subsection{Black Hole Information}
The black hole information paradox is perhaps the best example of the failure of the effective field theory of gravity.  We will not review the details of the argument, as a nice overview can be found in \cite{ArkaniHamed:2007ky}.  The most convincing form of the paradox is the nice slice argument \cite{Lowe:1995ac}.  One can pick a class of `nice' spatial slices along which the curvature is small and compute the entanglement entropy between the inside and outside of the black hole along these slices.  Before the curvature of the slices becomes large, one finds the entanglement has grown beyond that which is consistent with unitary evolution.

It has been suggested by many authors that a breakdown in locality could resolve the problem, although the origin of this non-locality is varied \cite{Lowe:1995ac,Susskind:1993if,ArkaniHamed:2007ky}.  In \cite{ArkaniHamed:2007ky}, it was suggested that the imprecision of local operators could be the source of this non-locality.  We will review their argument using the limits of locality from the worldline operators constructed above.

The nice slices that are used to construct this paradox have the feature that, inside the black hole, the proper time between the slices becomes very small.  However, when these time intervals become too small, it is not possible to distinguish the different slices with gauge invariant observables.  If one uses particles of mass $m$ to probe the slices, then the best one can resolve each slice is $\delta t > m^{-1} > M_{Pl}^{-1}$.  Because there is a finite amount of proper time $\tau_{in}$ inside the black hole, we can only resolve a finite number of slices, namely $N_{max} =\frac{\tau_{in}}{\delta t} = R_{S}M_{Pl}$ where $R_{S}$ is the Schwarzschild radius.  In order to resolve the Hawking radiation, the time steps outside the black hole can be as large as the wavelength $\delta t_{out} < R_{S}$.  Thus the total proper time outside that can be described is $\tau < N_{max} \delta t_{out} < R_{S}^2 M_{Pl}$. 

While this does not resolve the problem, it suggests that a complete, unitary description may still be compatible with effective field theory \footnote{This is compatible with \cite{Hayden:2007cs} because it is required that the entanglement is measured from the time of black hole formation.}.  In particular, the maximum timescale we can describe with our observables is shorter than the time necessary to demonstrate information loss $\sim M_{Pl}^{2} R_{S}^3$.  Note that our timescale is shorter than that of \cite{ArkaniHamed:2007ky} because our resolution was set by $M_{Pl}$ rather than the mass of the black hole.  This is consistent with a perturbative failure of locality rather than a failure arising from $e^{-S}$ effects.  Let us remind the reader that this does not imply a fundamental breakdown in locality.  Specifically, the S-matrix may still be local to all orders in perturbation theory.  Nevertheless, it should be stressed that one cannot rule out the existence of operators whose non-local corrections are non-perturbative.  We do not expect this is the case given that corrections to local operators in gauge theories arise perturbatively.

\subsection{Locality in de Sitter}

Classically, de Sitter space is one of the simplest solutions to Einstein's equations.  Yet, when it comes to formulating a quantum theory, de Sitter is notoriously confusing \cite{Witten:2001kn}.  This has suggested to some that de Sitter will break down in some finite time \cite{Polyakov:2007mm, Tsamis:1996qq,Dyson:2002pf, Goheer:2002vf}, likely shorter than the Poincare recurrence time \cite{Banks:2002wr,Page:2006dt}.  The fact that string compactifications seem only to give meta-stable de Sitter solutions \cite{Silverstein:2001xn, Kachru:2003aw} has been taken as further evidence in favor of this view.

While it is widely believed that a quantum theory of de Sitter should breakdown, the timescale on which this occurs has been the subject of some debate.  In \cite{ArkaniHamed:2007ky}, it was suggested that non-locality could somehow clarify the issues in the same way it did for the black hole.  With this in mind, let us examine our proposed operators in de Sitter backgrounds.  Of course, any large non-local corrections will signal a breakdown of the locality of our observables but not necessarily an instability of de Sitter.

We begin by discussing observables localized on timelike worldlines.  These observables are interesting for two reasons.  Firstly, timelike worldlines have often been used to define measures and classically measurable quantities in de Sitter space or eternal inflation (e.g. \cite{Bousso:2006ev,Vanchurin:2006qp}).  Secondly, these observables are more amenable to direct computation.

We will work on the static patch where the metric is given by
\beq
ds^2 = -(1-\frac{r^2}{l^2})dt^2+(1-\frac{r^2}{l^2})^{-1}dr^2+ r^2 d\Omega^2,
\eeq
with $l = \sqrt{\frac{3}{\Lambda}}= H^{-1}$.  We will be interested in observables that would be localized in the $m \to \infty$ limit on the timelike geodesic that sits at $r=0$.  At finite $m$, we would like to know over what time scale the geodesic approximation is valid.  Let us expand around the solution $\bar{X}^{0}= t$ and $\bar{X}^{1}\equiv \bar{R} =0$, focusing on the fluctuations in $R$.  In order to have gauge invariant observables we should extend the worldlines to the asymptotic past and future, but we will start by working on some finite time interval.

We will expand the action for a massive particle \footnote{We have integrated out $e$ to get the Nambo-Goto form of the action for a massive particle} around $R=0$,
\beq
S = m \int_{0}^{1} d\tau \sqrt{ -(1-\frac{R^2}{l^2})(\frac{\partial t}{\partial \tau})^2+(1-\frac{R^2}{l^2})^{-1}(\frac{\partial R}{\partial \tau})^2+\ldots}.
\eeq
When $\frac{\partial t}{\partial \tau} \gg \frac{\partial R}{\partial \tau}$, and $R \ll l$ we get
\beq
S \simeq \frac{i m}{2}\int_{0}^{T} dt (\frac{\partial R}{\partial t})^2+ \frac{R^2}{l^2}(1+(\frac{\partial R}{\partial t})^2)+\ldots .
\eeq
where $T= \bar{X}^{0}(1)$.  This is an inverse harmonic oscillator with $\omega^2 \sim - H^2$.  This unstable potential will likely make our geodesic approximation break down in finite time.  In order to estimate the time scale, we will work with a Wick rotated $(X^{0} \to i X^{0})$, positive definite metric.  Then, our action is
\beq
-S_{Euclid} \simeq -\frac{m}{2}\int_{0}^{T} dt (\frac{\partial R}{\partial t})^2 - \frac{R^2}{l^2}(1-(\frac{\partial R}{\partial t})^2)+\ldots
\eeq
We would like to estimate the period over which the integral will be controlled by the geodesic.  This model can be solved exactly (see \cite{Albrecht:1992kf} for a solution using squeezed states).  However, we will offer a more intuitive (but non-rigorous) argument that will be useful when discussing more general backgrounds.  Regardless, the exact solution yields the same answer in this case.

We are interested in paths that contribute significantly to the path integral.  From the kinetic piece, we can estimate that paths with maximum deviation $|R|_{max}=L$ contribute significantly to the path integral when $1 > S_{kinetic}(L) \geq \frac{m L^2}{2 T}$.

The kinetic term already seems problematic as $T \to \infty$ for finite $m$, but we will return to that concern later.  We would like to know over what timescale the path integral is controlled by the kinetic terms and not by the instability due to the inverted potential.  This can be estimated by 
\beq
\label{kinetic}
|S_{potential}(L)| \leq  m H^2 L^2 T  <  \frac{m L^2}{T} \leq S_{kinetic}(L)
\eeq
Thus, we expect our local description (for the timelike worldline operators) to become non-local for times $t> T_{crit} = H^{-1}$.  This is an extremely short timescale.  We will find better behaved observables, but it is nonetheless surprising that non-relativistic massive particles are not helpful for localization.

Although the worldline path integral is not easily computable in de Sitter space, the Green's functions for scalar fields are known \cite{Bousso:2001mw,Polyakov:2007mm}.  In soluble backgrounds like flat space, the worldline path integral computes the propagator for a scalar field \cite{Cohen:1985sm,Louko:2000tp}.  Similarly, one expects that the Green's function in de Sitter space is the result of the path integral for a single particle in the background.  For our purposes, we only need to assume that the Green's functions give the same qualitative features as the worldline path integral.  Under this assumption, (\ref{ob2}) takes the form
\beq
\label{green}
\mathcal{O}_{\Psi_{i=1..N}} = \mathcal{N} \int d^d x \sqrt{-g} \mathcal{O}(x) \prod_{i=1}^{N} G(x,y_{i})+\mathcal{O}(M_{Pl}^{-2})
\eeq
where $G(x,y)$ is the Feynman propagator for a massless (or massive) scalar field.  We will now use this to estimate the behaviour of our observables for relativistic particles (both massless and massive).  These observables would be exactly local in the $\alpha' \to 0$ limit of (\ref{ob2}).  

For a miminally coupled {\it massless} scalar, the Green's function takes the form \cite{Tolley:2001gg}
\beq
\label{massless}
G(x,x') = \frac{H^2}{4 \pi^2} ((1-P)^{-1}+ \ln(1-P)+\ln(2)-\frac{1}{2}),
\eeq
where $P$ is related to the geodesic distance $\ell(x,x')$ by $P=\cos \ell$ (imaginary $\ell$ is timelike).  The pole at $P=1$ is on the lightcone, and is what we used to get localization in flat space.  However, this propagator has two types of infrared (IR) problems.  First of all, the constant term means that the propagator does not vanish at large spatial separations.  Because we will have to integrate over one of the endpoints, this will always give a contribution to (\ref{green}) of the form
\beq
\mathcal{O}_{\Psi_{i=1..N}} \sim \int dx \sqrt{-g} \mathcal{O}(x)+\ldots .
\eeq
The second problem is that the $\ln(1-P)$ is actually divergent in the IR ($|P| \to \infty$).  We could have come to the same conclusion from knowledge of inflationary pertubations.  There one is calculating a constant time correlation function $\langle \phi(x,t) \phi(y,t) \rangle$, which is the same as our propagator for space-like separation.  The IR divergence can be seen by integrating the $k$ independent power spectrum to get
\beq
\langle \phi(x,t) \phi(y,t) \rangle \simeq -\frac{H^2}{4\pi^2} \ln\frac{|x-y|}{L}.
\eeq
Both of these features are special to de Sitter or quasi-de Sitter backgrounds, and are the primary obstacle to the quasi-locality of these operators.

The Green's function in $d$ dimensions for a {\it massive} scalar in the Bunch-Davies vacuum is given by \cite{Chernikov:1968zm, Tsamis:1992xa, Spradlin:2001pw}
\beq
G(x,x') = \frac{\Gamma(h_{+}) \Gamma(h_{-})}{(4 \pi)^{\frac{d}{2}} \Gamma(\frac{d}{2})} F(h_{+},h_{-}; \frac{d}{2}; \frac{1+P(x,x')}{2}),
\eeq
where $F$ is a hypergeometric function and $h_{\pm} \equiv \frac{d-1}{2} \pm \sqrt{(\frac{d-1}{2})^2- l^2 m^2}$.  The first important fact about the propagator is that it has a pole at $P=1$.  This offers hope that localization may be possible for some observables \footnote{The Green's function for the $\alpha$ vacua has an additional pole on the antipodal point.  This would seem to cause localization to fail immediately for $e^{\alpha} \neq 0$.}.

In an attempt to localize operators, we will use (\ref{ob2}) with massive particles.  These observables may localize on the intersection of lightcones if we can avoid the IR problems of the massless observables.  For large $P$, the propagator behaves as
\beq
G(x,x')  \to C_{1} P^{-h_{+}}+C_{2}P^{-h_{-}}.
\eeq
From the form of $h_{\pm}$, if $m^2 \gg H^2$ these contributions will be suppressed on large distances.  Therefore, in order to have localized observables, we need a sufficient hierarchy between $M_{Pl}$ and $H$ such that
\beq
M_{Pl} \gg m \gg H.
\eeq
Under these circumstances, the non-gravitational contribution yields operators localized at the intersection of null geodesics originating at $\mathcal{I}_{\pm}$.  If we only have asymptotic de Sitter in the past or future, the amount of local information is limited.  Even classically, if we only have null geodesics originating at $\mathcal{I}_{-}$ ($\mathcal{I}_{+}$), one cannot arrange for the geodesics to intersect arbitrarily far in the future (past).  At best one could have local information `near' $\mathcal{I}_{-}$ ($\mathcal{I}_{+}$).

However, until we understand the gravitational fluctuations in de Sitter, we cannot guarantee local behavior.  At the very least, one would like to calculate the perturbative corrections analogous to (\ref{corr2}).  However, the form of the graviton propagator in de Sitter will cause these corrections will be large.  The graviton propagator for a locally \footnote{Fully de Sitter invariant propagators have many unphysical properties \cite{Woodard:2004ut} in addition to the same IR divergent behavior as the massless scalar.} de Sitter background can be built from the massless scalar propagators \cite{Woodard:2004ut,Tsamis:2005je,Janssen:2007ht}.  In order to get small gravitational corrections, one needs the propagator to fall off at large distances. In particular, based on (\ref{corr}) and (\ref{green}), we expect corrections of the form
\beq
\mathcal{O}(M_{Pl}^{-2}) \sim \frac{\langle \mathcal{O}(\tilde{x}) \rangle}{M_{Pl}^{2}} \sum_{a,b} \int d \tau_{a} \int d \tau_{b} p^{\mu}_{a}(\tau_{a}) p^{\nu}_{a}(\tau_{a}) \mathcal{P}_{\mu \nu \lambda \rho}(X_{a},X_{b}) p^{\lambda}_{b}(\tau_{b}) p^{\lambda}_{b}(\tau_{b}),
\eeq
where $p_{a}^{\mu}$ is the four-momentum of a worldline and $\mathcal{P}_{\mu \nu \lambda \rho}$ is the graviton propagator in de Sitter.  Thus, the IR divergent terms in (\ref{massless}) will produce large gravitational corrections to our operators.  These corrections prevent us from getting quasi-local behaviour for any of our observables.  The meaning of the IR divergences in de Sitter is the subject of much debate \cite{Tsamis:1992xa,Tsamis:2007is,Garriga:2007zk}.  It remains an open question if one can formulate gravitational perturbation theory in de Sitter in a sensible way \cite{Janssen:2007ht,Einhorn:2002nu,Losic:2006ht}.

In gauge theory, similar divergent contributions from the Wilson line are the signal of confinement.  This suggests a possibly interesting (but very speculative) interpretation of de Sitter as a `confining' phase for local observables.

\subsection{General Backgrounds}

In order to understand more general backgrounds, it is worth recalling why our different examples have succeeded and failed.  We demonstrated explicitly that in flat space with perturbative gravity one can construct quasi-local gauge invariant observables.  Another canonical example is Anti-de Sitter (AdS) which has been studied extensively in the context of AdS-CFT.  Worldlines in AdS play an important role as they can be related to correlation functions in the CFT.  The geodesic approximation can be shown to give the leading behavior \cite{Kraus:2002iv}, and subleading corrections can be computed \cite{Fidkowski:2003nf}.  Attempts have also been made to give a precise definition of local bulk operators in terms of the CFT correlators \cite{Hamilton:2005ju}.

The success of the asymptotically flat and AdS examples can be attributed to poles in the worldline path integral (and Green's function), that appear on the light cone.  These poles are related to the fact that points on the light cone are separated by zero proper time.  When several worldlines are used together, the observables get localized near the intersection of null worldlines.

In any background, one expects the Green's functions to have poles on the light cone.  This does not depend on the global structure or particular details of the background.  They are simply a result of working with a Lorentzian geometry with null geodesics.  Therefore, this basic ingredient that was helpful in calculable backgrounds should be available in any background.

When discussing de Sitter space our observables failed to be local.  As we demonstrated, the poles in the propagator are not sufficient to guarantee localization.  The obstruction in that case was divergent behavior of the Green's functions in the IR.  Global features of the background determine the behavior of the propagator as one moves to infinite separation.  If the worldline and/or graviton propagator does not approach zero quickly enough at large seperation, then the integral over the spacetime volume will get large contributions away from the poles.

There is reason that one might hope that such IR problems are special to de Sitter and quasi-de Sitter backgrounds.  Specifically, they seem to be related to large fluctuations of the metric and light fields outside the horizon.  More general backgrounds with asymptotic behavior that is not dS like will hopefully not suffer from the same difficulties.  In such cases, one should be able to construct quasi-local observables.  Therefore, we expect that quasi-local observables exist for many types of asymptotic behavior.

Finally, there is one concern that may apply to these observables.  We have assumed in all cases that the stationary phase approximation holds in then context of interest.  There may be circumstances where corrections to it become important and obstruct the construction of quasi-local operators.  For more discussion of these points, see \cite{Louko:2000tp}

\section{Conclusions}

Motivated by the analogy with gauge theory, we studied the non-local gravitational corrections to local operators in backgrounds with asymptotic behavior.  We proposed a class of manifestly gauge invariant operators in (\ref{ob2}).  These operators are exactly local in the absence of dynamical gravity, but receive perturbative, non-local corrections when gravity is included.  This type of behavior is consistent with our expectations from gauge theory.  Using these observables, we studied apparent breakdowns in locality in both black hole and de Sitter space backgrounds.  In the case of de Sitter space, our observables failed to give quasi-local behavior.

For a general background, we found that observables built from the worldlines of non-relativistic particles, fail to give quasi-local behavior in many backgrounds.  Those built from null worldlines were seemingly better behaved and could potentially be useful in many settings.

One goal of this work was to understand the timescales over which de Sitter space can be consistently defined.  There are many reasons to believe that de Sitter is unstable \cite{Polyakov:2007mm, Tsamis:1996qq, Dyson:2002pf, Goheer:2002vf,Page:2006dt,Silverstein:2001xn, Kachru:2003aw}, but the timescale on which this must occur is unclear.  It was suggested in \cite{ArkaniHamed:2007ky} that the non-locality of gravity may play a role in a proper understanding of de Sitter.  The observables we studied failed to be quasi-local in de Sitter space due to the IR divergences in the propagators of massless fields.  By analogy with confinement, this may suggest an interpretation of the IR divergences as a failure of locality.  

\acknowledgments
I would to thank Peter Graham, Matt Headrick, Michael Mulligan, Eva Silverstein, Dusan Simic, David Starr, Bill Unruh and Sho Yaida for helpful discussions.  This project was supported in part by NSERC, the Mellam Family
Foundation, the DOE under contract DE-AC03-76SF00515 and the NSF under contract 9870115.

\bibliography{observbib}
\end{document}